\documentclass[cits]{PoS}
\usepackage{graphicx} \usepackage{natbib} 
\title{The Resolved Near-Infrared Extragalactic Background}

\ShortTitle{The Resolved Near-Infrared Extragalactic Background}

\author{\speaker{Ryan Keenan}%
      \thanks{Co-authors: A.~J.~Barger, L.~L.~Cowie, W.-H.~Wang, ApJ 723:40-46, 2010}\\
      University of Wisconsin, Madison\\
       E-mail: \email{keenan@astro.wisc.edu}}

\abstract{We present a current best estimate of the integrated
  near-infrared (NIR) extragalactic background light (EBL)
attributable to resolved galaxies in $J$, $H$, and $K_s$.  Our results for
measurements of $\nu I\nu$ in units of nW~m$^{-2}$~sr$^{-1}$ are
$11.7^{+5.6}_{-2.6}$ in $J$, $11.5^{+4.5}_{-1.5}$ in $H$ and
$10.0^{+2.8}_{-0.8}$ in $K_s$.  We derive these new
limits by combining our deep wide-field NIR photometry from five widely separated fields with other studies from the literature to create a galaxy counts sample that is highly complete
and has good counting statistics out to $JHK_s \sim 27-28$.  As part of this
effort we present new ultradeep $K_s-$band galaxy counts from 22 hours of
observations with the Multi Object Infrared Camera and Spectrograph
(MOIRCS) instrument on the Subaru Telescope. We use this MOIRCS $K_s-$band
mosaic to estimate the total missing flux from sources beyond our detection
limits.   Our new limits
to the NIR EBL are in basic agreement with, but $10-20$\% higher than
previous estimates, bringing them into better agreement with estimates of the
total NIR EBL (resolved + unresolved sources)
obtained from TeV $\gamma-$ray opacity measurements and recent direct
measurements of the total NIR EBL, as well as recent model estimates for the total light from galaxies.  We examine field to field variations in our photometry to show that the integrated light from galaxies is isotropic to within uncertainties, consistent with the expected large-scale isotropy of the EBL.}

\FullConference{Cosmic Radiation Fields: Sources in the early Universe\\
                November 9-12, 2010\\
                Desy, Germany}

\begin{document}

\section{Introduction}

The near-infrared (NIR) extragalactic background light (EBL) is the total light from resolved and
unresolved extragalactic sources in the NIR.  This represents the
integrated light from all star and galaxy formation processes over the history of the
universe that has been redshifted into the NIR.  Some fraction of the NIR EBL
can be resolved as the light from individual galaxies (Integrated Galaxy
Light, IGL), but the fractional
contribution from unresolved (and perhaps unresolvable) sources is not well
constrained.  A measurement of the total NIR EBL minus the IGL
will provide insight into the energy budget of the early universe.

While direct unresolved measurement of the NIR EBL is technically difficult due
to complex foregrounds, several authors have reported a measurement of the
total (resolved plus unresolved) NIR EBL
\citep{Haus98,Dwek98,Gorj00,Wrig01,Mats01,Mats05,Camb01,Leve07} and found it to be a factor of two
or more above the IGL obtained through source
counts \citep{Mada00,Tota01,Thom03}.
This is known as the NIR background excess (NIRBE).  The spectral energy distribution
(SED) of this measured excess, in some cases, appears very similar to that of zodiacal light,
which suggests there may be a foreground contamination issue.  Another
possible solution to this excess is a large population of undetected faint
galaxies and/or population III (PopIII) stars contributing a large fraction of
the NIR background  (see \citep{Kash05a,Haus01} for reviews).

Since the NIR EBL presents a source of opacity for TeV $\gamma-$rays via
pair production, the density of the background light can, in principle, be
measured via direct observation of TeV blazars.  This method relies on
assumptions of the intrinsic blazar spectrum and the SED of the EBL from
ultraviolet to the NIR, both
of which are poorly observationally constrained.  Nevertheless, TeV $\gamma-$rays provide an
independent estimate of the NIR EBL that can be used to help determine what
fraction of the background light can be attributed to resolved sources, and
how much may arise from faint and possibly exotic sources in the early
universe.

\begin{figure}
\begin{center}
\includegraphics[width=120mm]{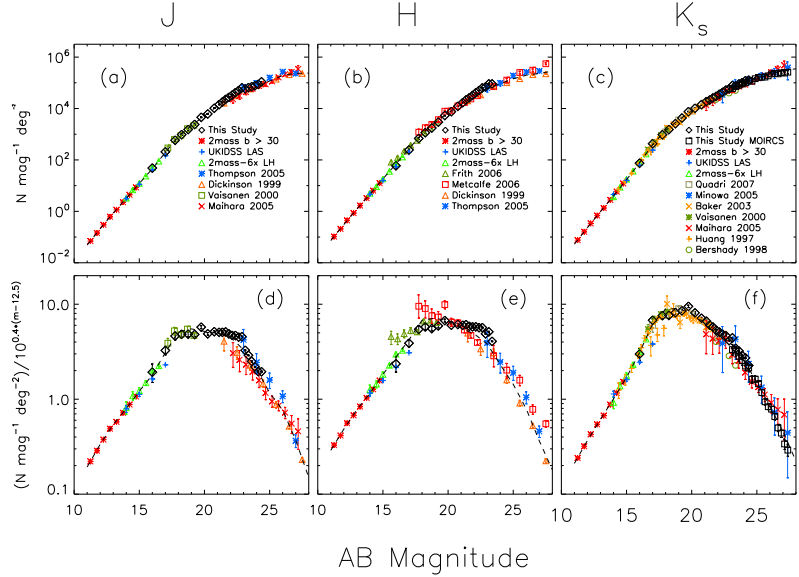}
\caption{\label{mycounts} (a-c)   Galaxy counts as a function of apparent
  magnitude.  Average galaxy counts from our deep,
  wide-field data on 5 fields (this study) are denoted by black diamonds.  The $K_s-$band counts from our Subaru MOIRCS data
  (this study MOIRCS) are denoted by black squares.  Error bars for this work are approximately the
  size of the plot symbols.  The counts determined by \citep{Keen10} from the
  2MASS \citep{Skru06} field with Galactic latitudes of $|b|>30$ are denoted by red
  asterisks. The counts determined by \citep{Keen10} from the the 2MASS-6x
  Lockman Hole survey \citep{Beic03} are denoted by green triangles.  The counts determined
  by \citep{Keen10} from three subfields of the UKIDSS \citep{Lawr07} LAS are denoted by blue
  plus symbols. Other data points are
  taken from the studies listed in the plots.  The dashed curve shows our
  error-weighted least squares running average (described in
  Section~2) from which we calculate the NIR IGL.  (d-f) The
  same data as in (a-c) but divided by an arbitrarily normalized Euclidean
  model with slope $\alpha = 0.4$. }
\end{center}
\end{figure} 

\section{NIR Background Due to Resolved Galaxies}
\label{nir_resolved}
\begin{figure}
\begin{center}
\includegraphics[width=90mm]{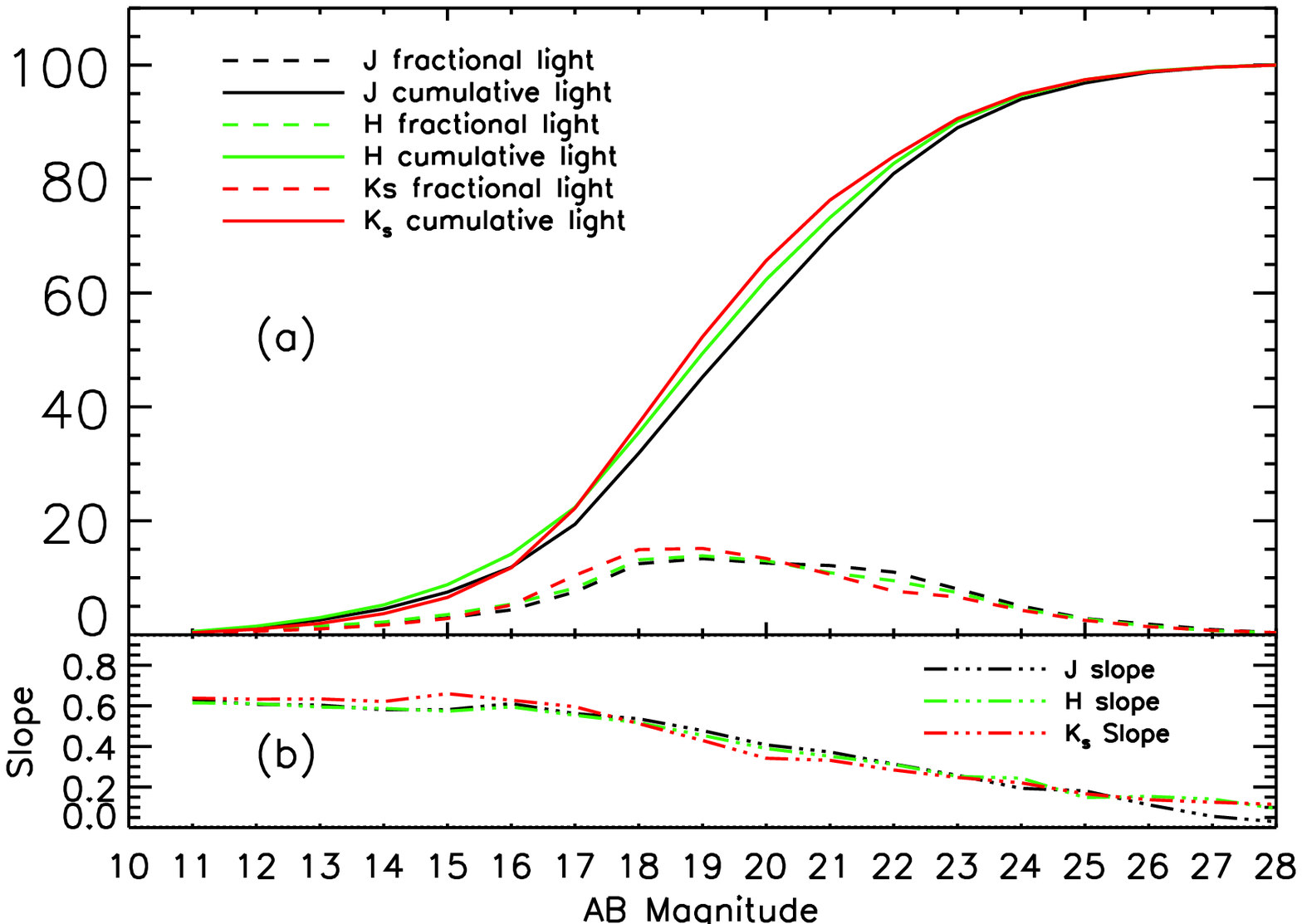}
\caption{\label{nirebl_calc}(a) Percent contribution to the IGL as a function
  of apparent magnitude.  Solid curves show the cumulative contribution
  from galaxies to the NIR IGL as a function of apparent magnitude.  Dashed
  curves show the fractional contribution per each magnitude bin. This
  demonstrates that the vast majority of the IGL ($\sim90$\%) arises from
  galaxies in the apparent magnitude range $\sim 15 < JHK_s < 24$ where this
  study is highly complete and has good counting statistics.  (b)  Slope of
  the galaxy counts curve as a function of apparent magnitude.
  Dash-dotted curves show the measured slope of the galaxy counts curve as a
  function of apparent magnitude.  When the slope drops below $\alpha = 0.4$
  the total light from galaxies becomes convergent.  }
\end{center}
\end{figure} 
The observations and data reduction for the NIR survey data presented in this paper are described in detail in \citep{Keen10}, \citep{Wang09} and \citep{Barg08}. In Figure~\ref{mycounts} we show our completeness-corrected and averaged
galaxy counts with those drawn from the literature.  However, before combining
all of our data in this way, we first investigated field to field variations for the IGL for
our five fields over the magnitude range $14.5 < JHK_s < 22.5$ where all five
are highly complete and have good counting statistics.  We found the
IGL over this magnitude range to be consistent across the four non-cluster
fields (CLASXS, CLANS, CDF-N and SSA13 $\sim 7-8$
nW~m$^{-2}$~sr$^{-1}$) with a $1~\sigma$
dispersion of $\pm 0.5$
nW~m$^{-2}$~sr$^{-1}$.  As such, we find that the IGL is consistent with large-scale istropy, an expected signature of the EBL (see \citep{Kash05a,Haus01} for reviews).

In the A370 cluster field, our IGL measurements were
$\sim 2-3$ nW~m$^{-2}$~sr$^{-1}$ higher in all bands with the peak
contribution to the excess light arising from galaxies at $JHK_s \sim 17$,
consistent with an excess of $L_*$ galaxies at a redshift of $\sim 0.4$ as in
A370.  We include the cluster field in our average counts for this study
because in a survey of a few square degrees such as this, roughly one massive
cluster will be present.

Figures~\ref{mycounts}(d-f) display the same data as in
panels (a-c) after dividing by an arbitrarily normalized
Euclidean model (of $\alpha = 0.4$ in the form $N(m) = A \times$ 10$^{\alpha m}$, and $A$ is a
constant) to expand the ordinate and demonstrate where resolved galaxies
contribute the most to the IGL.  A flat line in (d-f) would imply an equal
contribution to the IGL at all magnitudes.  The areas of positive slope
show where galaxies contribute a larger fraction to the IGL as one moves
toward fainter magnitude.  The steep negative slope beyond $JHK_s > 23-24$
demonstrates the diminishing contribution of resolved galaxies to
the IGL at the faintest magnitudes.

The results of our NIR IGL calculation are shown in Figure~\ref{nirebl_calc}.  
Figure~\ref{nirebl_calc}b shows the calculated slope of the galaxy counts
curve as a function of apparent magnitude.  Due to simple geometrical effects,
when the slope drops below $\alpha = 0.4$ the total light from galaxies
begins to converge.  Figure~\ref{nirebl_calc}a shows that the vast majority ($\sim 90\%$) of the resolved IGL
arises from galaxies in the range $\sim 15 < JHK_s < 24$.  Over this entire
range our study is highly complete and has good counting statistics.

Furthermore, from Figure~\ref{nirebl_calc} it can be seen that for sources
fainter than $JHK_s = 28$ to make any appreciable contribution to the IGL
the counts curve would have to rise dramatically over several magnitudes
beyond the limits of current surveys.  Such a rise would quickly result in
several (or more) galaxies per square arcsecond, rendering any counting exercise
impossible due to confusion.  Faint galaxies certainly exist beyond the limits
of the deepest NIR surveys, because at high redshifts the faintest apparent
magnitudes observed are only probing a few magnitudes fainter than $L_*$ down
the luminosity function.  It is unknown whether there is a steep upturn in the
luminosity function toward faint magnitudes, but if such an upturn exists and
faint galaxies contribute significantly to the NIR EBL, they would need to be
so numerous as to be unresolvable.
As such, it may be safe to say that the
resolvable portion of the NIR EBL (the IGL) has, for the most part, been measured and
that the most important contribution to the resolved portion comes from
galaxies in the magnitude range $\sim 15 < JHK_s < 24$, for which the deep
wide-field data presented here are optimized.   

\section{$K_s-$band Missing Flux}
\label{missingflux}

\begin{figure}
\begin{center}
\includegraphics[width=90mm]{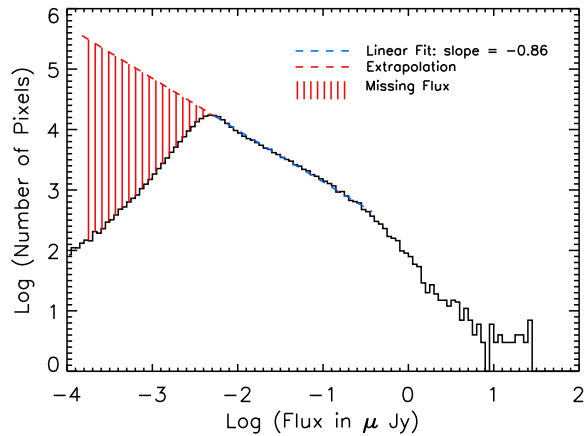}
\caption{\label{missingkflux} Histogram showing Log$_{10}$ of flux in $\mu$Jy
  vs. Log$_{10}$ of the number of pixels at that flux for all the pixels
  associated with galaxies in the Subaru MOIRCS $K_s-$band image.  The blue dashed line
  shows a linear fit from $\sim 0.005 - 0.4~\mu$Jy, which includes $\sim~60\%$
  of all pixels associated with galaxies and $\sim~99\%$ of such
  pixels containing fluxes higher than the peak of the histogram.  The slope
  of this line is -0.86.  We extrapolate the linear fit toward fainter fluxes to
the point where all the pixels in the image are accounted for (red dashed
line).The red hashed area shows the missing flux component corresponding
to 1.9 nW~m$^{-2}$~sr$^{-1}$ in the $K_s-$band.}
\end{center}
\end{figure}

\citep{Thom07a} estimated the missing flux component in the $J$ and $H-$bands
from the faint outer parts of galaxies and from galaxies below their detection
limits using a histogram of flux in all pixels associated with detected
objects.  We use a similar method to estimate the flux missed in the
$K_s-$band.  In Figure~\ref{missingkflux} we show a histogram of number
of pixels versus flux for all pixels associated with detected galaxies (object pixels) in our MOIRCS
$K_s-$band mosaic.  Noting the linear trend for fluxes $\sim
0.005 - 0.4~\mu$Jy, we fit a line to the data over this range (blue dashed
line). The portion of the histogram used in the fit represents 60\% of all
object pixels in the image and 99\% of object pixels with fluxes greater than
the turnover in the histogram at $\sim 0.005~\mu$Jy.  
The slope of the linear fit is -0.86.  \citep{Thom07a} find a slightly steeper
value of -1 by simply estimating the slope by eye.  Assuming that the true flux distribution for
faint pixels ($< 0.005~\mu$Jy) continues along the same trend, we extrapolate
the linear fit to approximate the shape of the histogram when all pixels in
the image are accounted for. Using this method, we calculate an estimate for
flux missed in the faint outer parts of galaxies and in galaxies that are
below our detection limits.  We find the missing flux component to be $\sim
22\%$ (1.9 nW~m$^{-2}$~sr$^{-1}$) of the total $K_s-$band light from resolved
galaxies. 

\section{Comparison With Previous Results}
\label{comparison}
\begin{figure}
\begin{center}
\includegraphics[width=90mm]{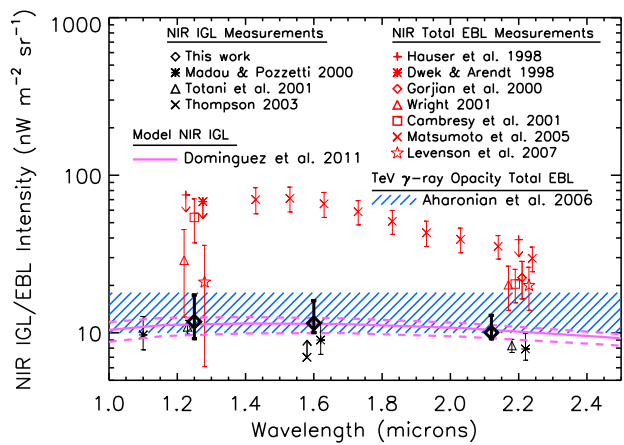}
\caption{\label{all_nirebl}NIR IGL and EBL measurements over the past decade
 as a function of wavelength ($\mu$m).  This work (black diamonds) is shown
 with a summary of measurements of the NIR IGL (other black symbols) via
  integrated galaxy counts \citep{Mada00,Tota01, Thom03}, model IGL (purple
  lines) \citep{Domi11}, total EBL (red
  symbols) via total light minus stars and zodiacal light \citep{Dwek98,
    Haus98, Gorj00, Wrig01, Camb01, Mats05, Leve07}, and via TeV $\gamma-$ray
  opacity measurements (blue hashed area) \citep{Ahar06}.  The range indicated
  for the $\gamma-$ray work shows the $1~\sigma$ error range.  Note that data
  points at $1.25, 1.6,$~and~$2.2~\mu$m have been shifted slightly in their
  abscissa values for clarity.  Arrows are used to denote upper and lower
  limits and otherwise error bars represent $1~\sigma$ confidence levels.
  Our results bring the measurement of the IGL into better
  agreement with TeV $\gamma-$ray observations and the most recent
  total NIR EBL measurements.  The lower limits on our data points show the
  $1~\sigma$ error estimates associated with the galaxy counts integration
  described in Section~2, while
  the upper limits show these same $1~\sigma$ error estimates plus the missing
  flux component derived in Section~3}
\end{center}
\end{figure}
 In Figure~\ref{all_nirebl} we show our IGL results alongside a summary of
measurements of the NIR IGL (black symbols) and EBL (red symbols) over the past decade.  The purple lines show the modeled IGL from galaxies at $z<4$ from \citep{Domi11}. The blue hashed area of Figure~\ref{all_nirebl} shows the allowed ($1~\sigma$) NIR EBL
intensity ($14 \pm~4$ nW m$^{-2}$ sr$^{-1}$) derived from High Energy Stereoscopic System (HESS) observations of
TeV blazars \citep{Ahar06}.   \citep{Mazi07} use 13 TeV blazars and a grid of
NIR background intensities to further constrain the NIR EBL and find
approximate agreement with the results of \citep{Ahar06}.  As noted earlier,
however, estimates of the NIR EBL from TeV $\gamma-$ray opacity measurements
rely on assumptions about the intrinsic SEDs of blazars and that of the EBL,
both of which are poorly constrained observationally.  Our results (black diamonds) are some $10-20$\% higher than
previous estimates of the IGL, which puts them closer to EBL estimates from
TeV blazar observations and the most recent total NIR EBL measurements.   

A large NIRBE has been found by several groups, with perhaps
the most striking result being that of \citep{Mats05}.  This excess, when
combined with other NIR and optical background measurements, showed an apparent
spectral break around 1~$\mu$m.  The excess was originally attributed to
PopIII stars, with the break corresponding to the redshifted Lyman limit for
these stars at $z\sim 10$.  However, a search for the possible absorption
imprint of this break on the $\gamma-$ray SED of blazars did not find evidence
for such a feature \citep{Dwek05a}.
 \section{Summary}
\label{summary}

Our new results for the NIR IGL
place the best current constraints on the total NIR light from resolved
galaxies and serve as a new lower limit to the total NIR EBL.   While these results are in relative agreement with previous
measurements, our numbers are $10-20$\% higher, bringing them into better
agreement with those derived from $\gamma-$ray experiments and the most recent
measurements of the total NIR EBL.  

We find the IGL to be roughly isotropic, consistent with the expectation of
large-scale isotropy in the EBL. We confirm that the starlight subtraction for
the most recent total NIR EBL measurements is correct (the reader is referred
to Keenan et al. \citep{Keen10a} for a detailed description of this measurement).

While our measurements cannot rule out the existence of a NIRBE due to PopIII
stars or other exotic early universe objects, our new lower
limits on the IGL and the upper limits
found from TeV $\gamma-$ray experiments \citep{Ahar06} could now be considered
in rough agreement with the most recent total NIR EBL
measurements in the $J-$band, and in near agreement in the $K-$band.

\end{document}